\documentclass[aps, prl, twocolumn, superscriptaddress, english]{revtex4}
\usepackage{graphicx}
\usepackage{cancel}
\usepackage{enumerate}
\usepackage[breaklinks=false, pdfborder={0 0 1}, backref=false, colorlinks=true, linkcolor=blue, citecolor=blue]{hyperref}
\usepackage{textcomp}
\usepackage{amsmath}
\usepackage{amssymb}
\usepackage{pgf}
\usepackage{soul}
\usepackage{subfigure}
\usepackage[english]{babel}
\usepackage[capitalise]{cleveref}
\usepackage{pdfpages}

\usepackage{tikz}
\usetikzlibrary{decorations.markings}
\usetikzlibrary{decorations.pathmorphing,patterns}
\usetikzlibrary{decorations}
\usetikzlibrary{calc}
\usetikzlibrary{intersections}
\usepackage{pgfplots}
\usepackage{mathtools}

\newcommand{\Trace}{{\mathrm{Tr}}}

\def\ud{\mathrm{d}}

\newcommand{\nep}{\textrm{e}}

\newcommand{\Hc}{\mathrm{H.c.} }
\renewcommand{\Im}{\mathrm{Im}}

\newcommand{\SIAM}{\scriptscriptstyle \mathrm{SIAM}}

\newcommand{\Tr}{{\rm Tr}}

\newcommand{\rmE}{\mathrm{E}}
\newcommand{\opa}[1]{{\hat{a}^{\phantom \dagger}}_{#1}}
\newcommand{\opadag}[1]{{\hat{a}^{\dagger}}_{#1}}

\newcommand{\opc}[1]{{\hat{c}^{\phantom \dagger}}_{#1}}
\newcommand{\opcdag}[1]{{\hat{c}^{\dagger}}_{#1}}
\newcommand{\opd}[1]{{\hat{d}^{\phantom \dagger}}_{#1}}
\newcommand{\opddag}[1]{{\hat{d}^{\dagger}}_{#1}}
\newcommand{\opfdag}[1]{{\hat{f}^{\dagger}}_{#1}}
\newcommand{\opf}[1]{{\hat{f}^{\phantom \dagger}}_{#1}}
\newcommand{\opn}[1]{{\hat{n}^{\phantom \dagger}}_{#1}}

\newcommand{\Ham}{\widehat{H}}

\newcommand{\sys}{\scriptscriptstyle \mathrm{S}}
\newcommand{\anc}{\scriptscriptstyle \mathrm{anc}}
\newcommand{\ret}{\scriptscriptstyle \mathrm{R}}

\begin{document}

\title{Efficient mapping for Anderson impurity problems with matrix product states}
\author{Lucas Kohn}
\affiliation{SISSA, Via Bonomea 265, I-34136 Trieste, Italy}
\author{Giuseppe E. Santoro}
\affiliation{SISSA, Via Bonomea 265, I-34136 Trieste, Italy}
\affiliation{International Centre for Theoretical Physics (ICTP), P.O.Box 586, I-34014 Trieste, Italy}
\affiliation{CNR-IOM Democritos National Simulation Center, Via Bonomea 265, I-34136 Trieste, Italy}

\begin{abstract}
We propose an efficient algorithm to numerically solve Anderson impurity problems using matrix product states. 
By introducing a modified chain mapping we obtain significantly lower entanglement, as compared to all previous attempts, while keeping the short-range nature of the couplings. 
Our approach naturally extends to finite temperatures, with applications to dynamical mean field theory, non-equilibrium dynamics and quantum transport.
\end{abstract}

\maketitle

\begin{figure}
\centering
\includegraphics[width=8cm]{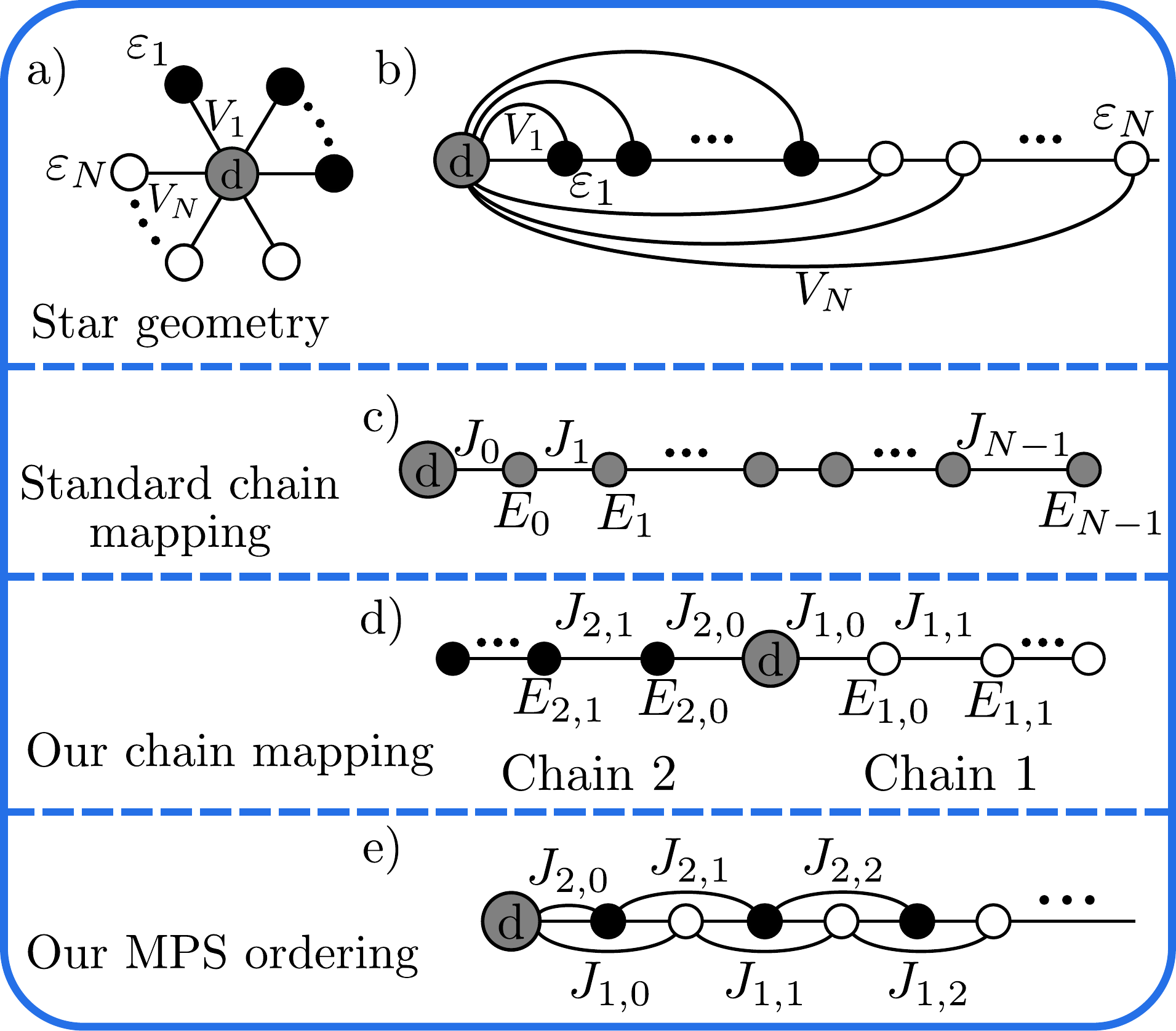}
\caption{
a,b) The star-geometry. a) The impurity orbital ``d'' hybridizes with a half-filled bath of conduction electrons. 
Here and below, gray circles denote partially filled orbitals, black/white fully-occupied/empty ones; 
b) The topologically equivalent chain with long-range hoppings, with orbitals ordered according to their increasing energy $\epsilon_k$; 
c) The conventional chain mapping, mixing all conduction states and leading to partly-filled orthogonal conduction orbitals;
d) Chain-mapping separating filled from empty conduction electrons;
e) Ordering of the sites used in our MPS. Notice that for $T=0$ we actually use two similar half-chains with spinless orbitals. 
}
\label{fig:star_geometry}
\end{figure}

{\em Introduction --- }
The single-impurity Anderson model (SIAM)~\cite{Anderson_PR61} is a cornerstone of condensed matter theory, both for the description of 
Kondo phenomena~\cite{Kondo_PTP64,Hewson_kondo:book}
and because it is a building block in Dynamical Mean Field Theory (DMFT) approaches to correlated materials~\cite{Metzner_PRL89,Georges_RMP96}. 
Several methods to calculate the impurity spectral function $\Im \, G (\omega)$ --- the crucial ingredient in the self-consistency DMFT loop~\cite{Georges_RMP96} --- 
have been developed, all of them with their own pros and cons. 
Continuous-time Monte Carlo~\cite{RMP_Gull_2011,PRB_Rubtsov_2005,PRL_Werner_2006} can deal with multiple band models, but can suffer from sign problems and 
requires analytic continuation to the real frequency axis. 
The Numerical Renormalization Group (NRG)~\cite{RMP_Bulla_2008, PRL_Stadler_2015,PRL_Bulla_1999, PRB_Zitko_2009, PRL_Zitko_2013} works directly on the real frequency axis, 
but is more difficult to extend to multiple orbitals and to finite temperature, although progress has been made~\cite{PRB_Bulla_2014}. 
Last, Matrix Product State (MPS) solvers based on Density Matrix Renormalization Group (DMRG)~\cite{White_PRL92} 
--- a very accurate approach for one-dimensional systems~\cite{Schollwoeck_RMP05,AoP_Schollwoeck_2011,montangero2018introduction} --- work at zero temperature, 
both on the imaginary~\cite{PRX_Wolf_2015,PRB_Linden_2020} and on the real~\cite{PRL_Hallberg_2004,PRB_Wolf_2014_Cheb,PRB_Verstraete_2015,Wolf_PRB14,PRX_Bauernfeind_2017} frequency axis,
although the latter suffers from growth of entanglement entropy during the time-evolution. 
Remarkably, the entanglement is even worse if a standard tight-binding mapping to a short-range model~\cite{Wilson_RMP75} is performed: the best approach 
works with long-range couplings~\cite{Wolf_PRB14}, in the so-called ``star geometry'' (see Fig.~\ref{fig:star_geometry}(a,b)).
Despite these continuous efforts, exact diagonalization --- requiring a truncation of the conduction electron degrees-of-freedom --- 
is still one of the most popular impurity solvers for DMFT~\cite{Georges_RMP96,PRB_Lu_2014}. 

The reason behind the failure of standard chain mappings in MPS-DMRG approaches is simple to understand:
The mapping involves an inevitable mixing of fully-occupied and empty conduction electron modes, below and above the Fermi energy, a mixing  
which is very detrimental to the entanglement in MPS-based dynamical calculation of the Green's function~\cite{Wolf_PRB14,PRB_Millis_2017}.

The present Letter presents an effective strategy to perform MPS-DMRG calculations based on short-ranged chain mappings that
{\em i)} show low entanglement and {\em ii)} can deal with finite temperature.
The crucial idea (sketched in Fig.~\ref{fig:star_geometry}(d)) is easy to understand at $T=0$: 
one needs to separate empty and filled conduction orbitals by constructing {\em two separate tight-binding chains}, 
independently coupled to the impurity orbital, an idea which has proven beneficial in Lanczos-based exact diagonalization already~\cite{PRB_Lu_2014}. 
To expand on that, allowing $T>0$ calculations, we borrow some of the tools developed in the field of open quantum systems with bosonic environments, 
notably the thermo-field transformation~\cite{ColPhen_Umezawa_1975,PRA_Vega_2015},
with fermionic applications to quantum transport~\cite{PRL_Weichselbaum_2018} and quench dynamics~\cite{PRB_Plenio_2020}.
The result is a very efficient MPS-DMRG impurity solver, with short-range couplings and low entanglement, which works also at finite $T$, 
and can be extended to deal with multiple orbitals \cite{PRX_Bauernfeind_2017} and out-of-equilibrium phenomena.

{\em Model and methods --- }
The SIAM~\cite{Anderson_PR61} consists of a single spin-full impurity orbital hybridizing with a bath of free conduction electrons:  
\begin{equation}
	\Ham_{\SIAM}(t) = \Ham_{\text{loc}}(t) + \Ham_{\text{hyb}} + \Ham_{\text{cond}} \;.
\end{equation}
The local Hamiltonian describes the impurity:
\begin{equation}
	\Ham_{\text{loc}}(t) = \sum_{\sigma} \varepsilon_d(t) \opddag{\sigma}\opd{\sigma} + U\, \opn{\uparrow}\opn{\downarrow} \;,
\end{equation}
where $\opddag{\sigma}$ creates an electron with spin $\sigma=\uparrow,\downarrow$ in the impurity orbital, $\opn{\sigma}= \opddag{\sigma}\opd{\sigma}$ is
the number operator, and $U$ the on-site Hubbard repulsion. 
Here we could allow for a time-dependent impurity level $\varepsilon_d(t)=E_d(t)-\mu$, where $\mu$ denotes the chemical potential of the conduction electron bath,
and even for a time-dependent $U(t)$, for possible applications to non-equilibrium problems.  
The impurity-bath hybridization is given by:
\begin{equation} \label{eq:AIM_hybr_discrete}
    \Ham_{\text{hyb}} = \sum_{\sigma} \sum_{k}  V_{k} \, \Big( \opddag{\sigma} \, \opc{k,\sigma} + \opcdag{k,\sigma} \, \opd{\sigma} \Big) \;,
\end{equation}
where $V_k$ is a (real) hybridization matrix element and $\opcdag{k,\sigma}$ creates a conduction electron in an orbital labelled by $k$, 
and $\Ham_{\text{cond}} = \sum_{k,\sigma} \; \epsilon_k\, \opcdag{k,\sigma} \, \opc{k,\sigma}$.
%
Here $\epsilon_k$ denotes energies referred to the conduction electron chemical potential $\mu$.

Following Wilson's NRG approach idea~\cite{Wilson_RMP75}, it is tempting to recast this problem into a nearest-neighbour tight-binding form, to better apply MPS/DMRG algorithms. 
Indeed, the only electronic orbital that couples to the impurity is the combination $\opa{0,\sigma} =  J_0^{-1} \sum_{k} V_k \opc{k,\sigma}$, 
where $J_0 = \sqrt{\sum_{k} V_k^2}$ enforces the correct normalization. 
Hence the problem can be cast into a tridiagonal tight-binding form by standard techniques, such as linear discretization plus Lanczos tridiagonalization or 
orthogonal polynomials \cite{JMath_Chin_2010,PRL_Prior_2010,Gautschi_2004}, as sketched in Fig.~\ref{fig:star_geometry}(c). 
But this is, unfortunately, not a good route: as shown in Ref.~\cite{Wolf_PRB14}, the entanglement properties of such a tight-binding chain 
scale worse than the plain application of MPS/DMRG techniques to the so-called {\em star geometry}, 
where one orders the orbitals according to their energy $\epsilon_k$, from the most negative ($\epsilon_k<0$) occupied orbitals up to the $\epsilon_k>0$ unoccupied ones, 
allowing for the appropriate long-ranged hybridization couplings $V_k$ (see Fig.~\ref{fig:star_geometry}(a,b)).

{\em Our mapping ---} We start discussing the zero-temperature case, omitting the spin index for a while. 
Let us re-define the original fermions as $\opc{k}\to \opc{1k}$. 
Their filled Fermi-sea state will be denoted by $|\mathrm{FS}_1\rangle = \prod_{k} \Theta(-\epsilon_k)\opcdag{1k} |0\rangle$,
where $\Theta(\cdot)$ is the Heaviside theta-function. 
Following Takahashi and Umezawa~\cite{ColPhen_Umezawa_1975}, we introduce a second ancillary fermionic operator $\opc{2k}$, 
which {\em will not couple to the impurity}, and put in a {\em fiducial ancillary state} 
given by their filled Fermi sea $|\mathrm{FS}_2\rangle = \prod_{k} \Theta(-\epsilon_k) \opcdag{2k} |0\rangle$. 
Define now a new canonical set of fermionic operators as follows~\cite{ColPhen_Umezawa_1975,PRA_Vega_2015,PRB_Plenio_2020,PRL_Weichselbaum_2018}:
\begin{equation} \label{eq:direct_gen}
\left( \begin{array}{c} \opf{1k} \\ \opf{2k} \end{array} \right) =
\left( \begin{array}{rr}  \cos \theta_k  & - \sin \theta_k \\   \sin \theta_k & \cos \theta_k \end{array} \right) 
\left( \begin{array}{c} \opc{1k} \\ \opcdag{2k} \end{array} \right) \;. 
\end{equation}
While originally introduced to deal with finite temperature, this transformation will turn out extremely useful even at $T=0$. 
For $\cos \theta_k=\Theta(\epsilon_k)$ and $\sin \theta_k=\Theta(-\epsilon_k)$,
\cref{eq:direct_gen} is a compact way of writing a conditional (on $\epsilon_k$) intertwined particle-hole transformations ---
$\opf{1k}=\opc{1k}$ and $\opf{2k} = \opcdag{2k}$ for $\epsilon_k>0$, while
$\opf{1k}=-\opcdag{2k}$ and $\opf{2k} = \opc{1k}$ for $\epsilon_k<0$ ---
so that $\forall k$:
\begin{equation}
\opf{1k} |\mathrm{FS}_1\rangle \otimes |\mathrm{FS}_2\rangle = 0 \hspace{5mm} \mbox{and} \hspace{5mm} 
\opfdag{2k} |\mathrm{FS}_1\rangle \otimes |\mathrm{FS}_2\rangle = 0 \;,
\end{equation}
hence $|\mathrm{FS}_1\rangle \otimes |\mathrm{FS}_2\rangle$ 
is the {\em vacuum} of the $\opf{1k}$, denoted by $|\emptyset_1\rangle$, 
and the {\em fully filled state} of the $\opf{2k}$, denoted by $ |\mathrm{F}_2\rangle$.
So, Eq.~\eqref{eq:direct_gen}, with the $T=0$ choice $\cos \theta_k = \Theta(\epsilon_k)$ and $\sin \theta_k = \Theta(-\epsilon_k)$, 
implies $|\mathrm{FS}_1\rangle \otimes |\mathrm{FS}_2\rangle \equiv |\emptyset_1\rangle \otimes |\mathrm{F}_2\rangle$.

Now, we supplement our SIAM with an ancillary conduction term
$\sum_{k} \epsilon_k\,\opc{2k} \, \opcdag{2k}$, such that the full kinetic energy reads:
\begin{equation}
 \sum_{k} \epsilon_k\, \Big( \opcdag{1k} \, \opc{1k} + \opc{2k} \, \opcdag{2k} \Big)  = 
 \sum_k \epsilon_k  \Big( \opfdag{1k} \, \opf{1k} + \opfdag{2k} \, \opf{2k} \Big) \;.
 \nonumber
\end{equation}
As for the hybridization, 
we have: 
\begin{equation}
\sum_{k}  V_{k} \, \opddag{}\! \, \opc{1k}  = 
\sum_{k}  \Big( V_{1k} \, \opddag{}\! \, \opf{1k} + V_{2k} \, \opddag{}\! \, \opf{2k}  \Big) \;,
\end{equation}
with $V_{1k} = V_k \, \cos \theta_k$ and  $V_{2k} = V_k \, \sin \theta_k$. 
Hence the impurity hybridizes with $\epsilon_k>0$ $\opf{1k}$-modes and $\epsilon_k<0$ $\opf{2k}$-modes, corresponding to the original $\opc{1k}$ modes, neatly separating empty and filled
conduction states, a key to a low-entanglement chain mapping. 

Interestingly, this naturally allows to capture the $T>0$ physics. 
Indeed, the expectation value of the original fermions $\opcdag{1k}\opc{1k}$ on the pure state 
$|\emptyset_1\rangle \otimes |{\mathrm F}_2\rangle = |\emptyset_1, {\mathrm F}_2\rangle$ is:
\begin{equation}
\langle \emptyset_1, {\mathrm F}_2 |  \, \opcdag{1k} \, \opc{1k} | \emptyset_1, {\mathrm F}_2\rangle = \sin^2(\theta_k) \;. 
\end{equation}
Hence the pure state $|\emptyset_1, {\mathrm F}_2\rangle$ will effectively mimic the Fermi occupation factor for the conduction electrons
provided we take:~\cite{PRB_Plenio_2020}
\begin{equation}
\sin^2(\theta_k)  \equiv  
\frac{1}{\nep^{\beta\epsilon_k} + 1} \;.  
\end{equation}
Notice that now, for such a general $\theta_k$, $|\emptyset_1, {\mathrm F}_2\rangle\neq |\mathrm{FS}_1\rangle \otimes |\mathrm{FS}_2\rangle$:
the pure-state $|\emptyset_1, {\mathrm F}_2\rangle$ captures thermal expectation values for the $\opc{1k}$, with a highly non-trivial structure of the states 
for the original fermions. 

\begin{widetext}
We carry out separate chain mappings for empty and filled modes to preserve their product state property \cite{PRA_Vega_2015,PRL_Weichselbaum_2018}, 
with initial orbitals $\opa{i,0} =  J_{i,0}^{-1} \sum_{k} V_{i,k} \opf{ik}$ ($i=1,2$). After reinstalling spin indices, we end up with the final tight-binding Hamiltonian
\begin{eqnarray}  \label{eq:HamChain}
\Ham_{\SIAM} = \Ham_{\text{loc}}  +  \sum_{\sigma} \sum_{i=1}^2 \bigg( J_{i,0} \left(\opddag{\sigma} \, \opa{i,0,\sigma} + \Hc \right) 
+ \sum_{n=1}^{\infty}  \left( J_{i,n} \opadag{i,n,\sigma}\opa{i,n-1,\sigma} + \Hc \right)  
+ \sum_{n=0}^{\infty}  E_{i,n} \, \opadag{i,n,\sigma}\opa{i,n,\sigma} \bigg) \;. 
\end{eqnarray}
\end{widetext}
While the Hamiltonian in Eq.~\eqref{eq:HamChain} has perfect chain geometry, see \cref{fig:star_geometry}(d), 
we employed a slightly different ordering of the sites, sketched in \cref{fig:star_geometry}(e).
Similar ideas were recently used to study quantum transport~\cite{PRE_Poletti_2020} and 1D systems with periodic boundary conditions~\cite{PRR_Rizzi_2020,ArXiv_Rizzi_2020}.
This ordering significantly reduces the entanglement at $T>0$, but requires next-nearest neighbor couplings, which we deal with by using the 
time-dependent variational principle (TDVP)~\cite{SIAM_Lubich_2015,PRL_Haegeman_2011,PRB_Haegeman_2016,AoP_Paeckel_2019,Scipost_Bauernfeind_2020,PRA_Kohn_2020}. We chose 2-site TDVP and checked projection errors by comparison with different MPS orderings (see SM).
A more detailed analysis of these issues will be presented elsewhere.

In the following we are interested in the retarded Green's function
\begin{align} \label{eq:retGF}
	G^{\ret}_{\sigma}(t)& = -i\Theta(t)  \, \Trace_{\sys} \left( \{\opd{\sigma}(t),\opddag{\sigma}(0)\} \hat{\rho}_{\sys} \right) \;,
\end{align}
where $\opd{\sigma}(t)$ is the Heisenberg impurity operator, $\hat{\rho}_{\sys}$ the thermal density matrix of the system, and $\{ \cdot , \cdot \}$ denotes the anti-commutator. 
The Fourier transform of the $G^{\ret}_{\sigma}(t)$ provides the experimentally accessible spectral function 
$A_{\sigma}(\omega) = - \frac{1}{\pi}\text{Im}\,\int_{-\infty}^{\infty} \ud t \ \nep^{i\omega t} \, G^{\ret}_{\sigma}(t)$.

\begin{figure}
	\centering
	\includegraphics[width=8cm]{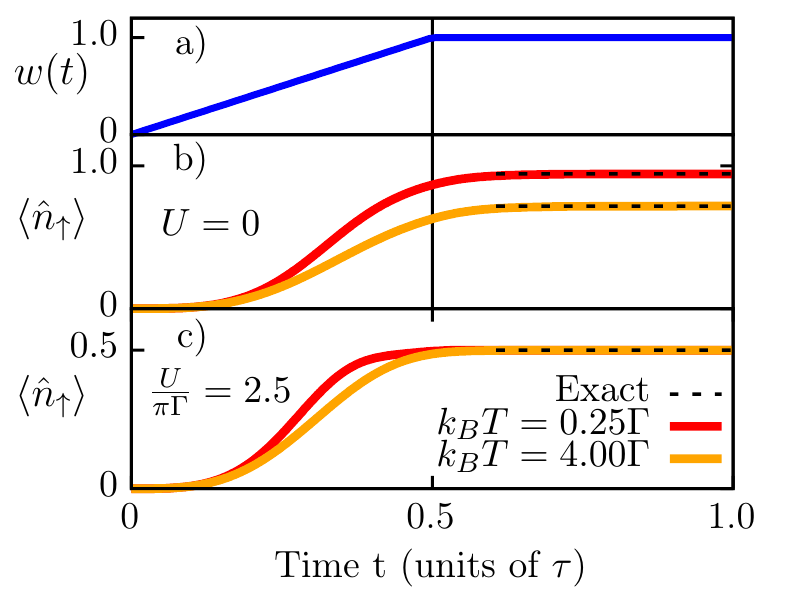}
	\caption{Top: Equilibration scheme employed to construct the thermal state of the system, see text for explanation.  
		Local and hybridization terms are ramped-up linearly and kept constant for $t/\tau>0.5$.
		Below: Evolution of $\langle \hat{n}_{\uparrow} \rangle$ for the noninteracting $U=0$ (b) and particle-hole symmetric case $U=2.5\pi\Gamma$ (c) at fixed impurity level energy $\varepsilon_d=-1.25\pi\Gamma$. Here $\tau=8\hbar/\Gamma$ and bond dimension $D=150$. 
		}
	\label{fig:n_eq}
\end{figure}

In general, $\hat{\rho}_{\sys}$ is the state where impurity and bath are in thermal equilibrium. 
Physically, equilibrium can be established by starting from a state where impurity and bath are isolated, by slowly turning on the hybridization between the two.  
At $T=0$, however, there is a short-cut, and $\hat{\rho}_{\sys}$ can be obtained by a DMRG ground state (GS) search~\cite{PRB_Verstraete_2015,Wolf_PRB14,PRX_Bauernfeind_2017}. 
For finite $T>0$ we cannot do that, and, following the previous idea of a real-time equilibration, the thermal state $\hat{\rho}_{\sys}$ is obtained through a TDVP evolution, 
working particularly well at higher temperatures, as detailed in the SM:
1) we initialize a state with empty impurity $|0\rangle$ and ``thermal'' conduction electrons, 
with a MPS pure state $|\psi_0\rangle=|0\rangle\otimes|\emptyset_1, {\mathrm F}_2\rangle$;
2) We evolve the system by slowly ramping-up the local and hybridization terms as 
$\Ham_{\SIAM}^{\text{eq}}(t)=w(t)(\Ham_{\text{loc}}+\Ham_{\text{hyb}})+\Ham_{\text{cond}}$, 
with $w(t)$ starting from zero and linearly approaching one, see \cref{fig:n_eq}(a), until reaching $\Ham_{\SIAM}$;
3) We finally relax the system with $\Ham_{\SIAM}$, getting to a final MPS pure state $|\rmE \rangle$ effectively encoding the correct equilibrium state,
such that $\hat{\rho}_{\sys} = \Tr_{\anc} \left(|\rmE \rangle\langle \rmE |\right)$.

To calculate $G^{\ret}_{\sigma}$ in \cref{eq:retGF} we first split it as $G^{\ret}_{\sigma}(t)=\Theta(t) \left( G^{>}_{\sigma}(t) - G^{<}_{\sigma}(t) \right)$, where 
$G^{>}_{\sigma}$ and $G^{<}_{\sigma}$ are the greater and lesser Green's functions, respectively, calculated in separate simulations. 
For the discussion we focus on $iG^{>}_{\sigma}(t) = \Tr_{\sys} (\opd{\sigma}(t)\opddag{\sigma}\hat{\rho}_{\sys})$, as $G^{<}_{\sigma}$ is similar. 
After writing explicitly the time dependence of the Heisenberg operators and reordering operators within the trace we get: 
\begin{equation}
iG^{>}_{\sigma}(t) = \Tr_{\sys+\anc} \left( \nep^{-i\hat{H}t/\hbar}\opddag{\sigma}\, |\rmE \rangle\langle \rmE| \,\nep^{i\hat{H}t/\hbar} \opd{\sigma}  \right) \;.
\end{equation}
Note that we traced over both the physical fermions and over the ancillary ones, as required by $\hat{\rho}_{\sys}$.
Defining states $|\phi^{>}_L(t)\rangle = \nep^{-i\hat{H}t/\hbar}\,\opddag{\sigma}\,|\rmE\rangle$ and $\langle\phi^{>}_R(t)| =  \langle \rmE|\,\nep^{i\hat{H}t/\hbar}\opd{\sigma}$, 
calculated by two independent TDVP evolutions, we finally rewrite 
$iG^{>}_{\sigma}(t)=\Tr(|\phi^{>}_L(t)\rangle\langle\phi^{>}_R(t)|)=\langle\phi^{>}_R(t)|\phi^{>}_L(t)\rangle$. 

\begin{figure}
	\centering
	\includegraphics[width=8cm]{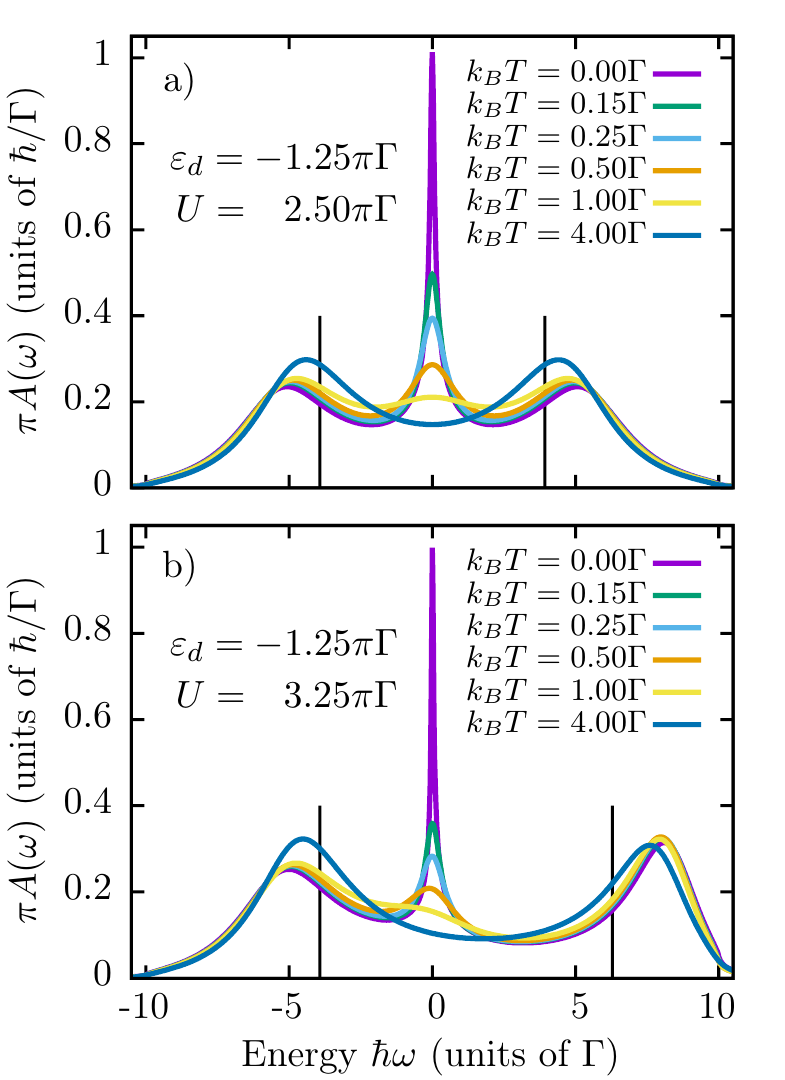}
	\caption{Spectral function as obtained from the Fourier transform of the retarded Green's function at different temperatures, showing the disappearance of the Kondo peak 
		for increasing $T$. The interaction is $U=2.5\pi\Gamma$ for the symmetric~(a) and $U=3.25\pi\Gamma$ for the non-symmetric case~(b). 
		Vertical lines mark impurity level energies at $\varepsilon_d$ and $\varepsilon_d+U$. In the symmetric case we can estimate a Kondo temperature $k_BT_{K}\sim 0.07 \Gamma$. 
		}
	\label{fig:A_omega}
\end{figure}

{\em Results ---}
In our simulations we used a semi-elliptical fermion bath hybridization with half bandwidth $W$ and coupling $\Gamma$, which in the continuum limit gives
$\sum_{k}V_{k}^{2}\delta(\epsilon-\epsilon_k)=\Gamma\sqrt{1-(\epsilon/W)^2}/\pi$. 
We fixed $W=10\Gamma$ and the impurity energy level $\varepsilon_d=-1.25\pi\Gamma$. From here on we use $\Gamma$ as our unit of energy.
As discussed, our method involves two steps: First, we determine the equilibrium pure state $|\rmE\rangle$ either by a direct DMRG GS-search ($T=0$) or by a dynamical ramp-up,
see \cref{fig:n_eq}(a), of $\Ham_{\text{loc}}$ and $\Ham_{\text{hyb}}$  ($T>0$). Second, we calculate the Green's function by TDVP evolutions starting from $|\rm E\rangle$.
 
To test the first (equilibration) step we start from the empty impurity state $|\psi_0\rangle=|0\rangle\otimes|\emptyset_1, {\mathrm F}_2\rangle$ and study the evolution of 
the spin-up impurity occupation $\langle \opn{\uparrow} \rangle$ and benchmark it against exact values,
for the noninteracting case ($U=0$), \cref{fig:n_eq}(b), where exact diagonalization is possible,  
and for the particle-hole symmetric case ($\varepsilon_d=-1.25\pi\Gamma$, $U=2.5\pi\Gamma$), \cref{fig:n_eq}(c), where the impurity is half-filled at any temperature 
$\langle \hat{n}_{\uparrow}\rangle=\langle \hat{n}_{\downarrow}\rangle=1/2$.
We find that $\langle \opn{\uparrow} \rangle$, starting from $0$, converges to the exact values at the end of the equilibration in all cases.  

In the second step we calculate the Green's function and --- by means of a Fourier transform --- the corresponding $A(\omega)$. 
To obtain a smooth $A(\omega)$ we employ ``linear prediction''~\cite{PRB_White_2008,PRB_Schollwoeck_2009,PRB_Verstraete_2015} to extrapolate simulation data to longer times. 
In the symmetric case, at sufficiently strong repulsion $U$, we find two symmetrically located broad peaks due to the impurity close to energies $\epsilon_d$ and $\epsilon_d+U$, 
and a Kondo peak at the Fermi energy, see \cref{fig:A_omega}(a)~\cite{PRB_Horvatic_1987}.
We observe the disappearance of the Kondo peak as $T$ increases. 
For the non-symmetric case of \cref{fig:A_omega}(b), at larger $U$, the Kondo peak becomes narrower but is still located at the Fermi energy.

\begin{figure}
	\centering
	\includegraphics[width=8.5cm]{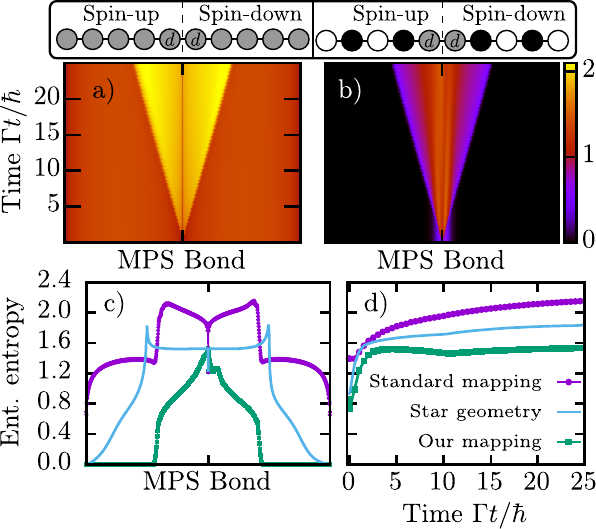}
	\caption{Top: Evolution of the entanglement entropy along the MPS for the original chain method (a) and our work (b) in the symmetric SIAM at T=0. 
	Bottom: (c) The entanglement entropy at time $t=25\hbar/\Gamma$ and (d) time-evolution of the maximum entropy 
	for the original chain mapping, \cref{fig:star_geometry}(c), our chain mapping, \cref{fig:star_geometry}(e), and the star geometry, \cref{fig:star_geometry}(b).}
	\label{fig:Entanglement}
\end{figure}

We now show how efficient is our mapping in terms of entanglement. 
We consider here $T=0$, where we use a modified MPS structure: in addition to the orderings in \cref{fig:star_geometry}(e), spin degrees of freedom are spatially separated 
--- which is known to be very efficient~\cite{PRB_Delft_2008,PRB_Verstraete_2015,PRL_Rams_2020} --- 
with spin-up on the left half of the MPS, spin-down on the right half, and the impurity in the middle. 
\cref{fig:Entanglement}(a,b) show that our mapping leads to significantly lower entanglement entropy with respect to a standard chain mapping,
with a characteristic ``light-cone'' spreading of entanglement, typical of short-ranged models. 
\cref{fig:Entanglement}(c) shows a snapshot of the entanglement entropy across the whole MPS at time $t=25 \hbar/\Gamma$: interestingly, not only we drastically improve on 
the standard chain mapping, but we also significantly improve on the star geometry (see SM for details). 
\cref{fig:Entanglement}(d) shows the time-evolution of the maximum entanglement in the MPS: the original chain mapping, and also the star geometry, seem to show a logarithmic growth of
entanglement, consistently with the analysis of Ref.~\cite{PRB_Millis_2017}, while the maximum entanglement entropy of our mapping seems to saturate, or in any case increase much more slowly.  
Finally, let us mention that our method shows excellent scaling with the bath size, as increasing the number of sites will only lead to an extension of the zero entanglement region (see \cref{fig:Entanglement}(b)) at the end of the chain, barely increasing the computational costs (see SM for details).

{\em Conclusions ---}  
We presented an efficient chain-mapping-based method to simulate Anderson impurity models using Matrix Product States. 
Our method overcomes a major problem of the original chain mapping, where mixing empty and filled sites led to large entanglement within the chain. 
Separating empty from filled sites and mapping them into separate chains, we drastically reduced the entanglement, providing significant performance improvements. 
In contrast to the star geometry, our method does not involve long-range couplings, but only next nearest neighbor terms. 
Using a thermo-field transformation the idea neatly generalizes to finite temperatures, where we demonstrated its capabilities by studying the Kondo physics regime. 
Future research directions include 
the application to non-equilibrium dynamics, and the implementation of the DMFT loop, where 
multi-orbital problems are within reach~\cite{PRB_Verstraete_2015,PRX_Bauernfeind_2017}.

We thank A. Amaricci, M. Capone, M. Dalmonte, M. Secl\`{\i} and E. Tosatti for discussions, and U. Schollw\"ock for a careful reading of the manuscript. 
Research was partly supported by EU Horizon 2020 under ERC-ULTRADISS, Grant Agreement No. 834402.
GES acknowledges that his research has been conducted within the framework of the Trieste Institute for Theoretical Quantum Technologies (TQT).
Simulations were performed using the ITensor library \cite{itensor}.

\nocite{Serafini_book_2017,PRB_Schroeder_2016,PRA_Singh_2010,PRB_Singh_2011,Scipost_Silvi_2019}


\end{document}


\title{Efficient mapping for Anderson impurity problems with matrix product states \\ Supplementary Material}
\author{Lucas Kohn}
\affiliation{SISSA, Via Bonomea 265, I-34136 Trieste, Italy}
\author{Giuseppe E. Santoro}
\affiliation{SISSA, Via Bonomea 265, I-34136 Trieste, Italy}
\affiliation{International Centre for Theoretical Physics (ICTP), P.O.Box 586, I-34014 Trieste, Italy}
\affiliation{CNR-IOM Democritos National Simulation Center, Via Bonomea 265, I-34136 Trieste, Italy}

\begin{abstract}
	This supplementary material deals with the issue of how to effectively prepare the initial equilibrium thermal state for the calculation of the Green's functions, and also gives details about the numerical ingredients used in our calculations. We investigate the performance of our chain-mapping-based method and compare its entanglement structure with that of the star geometry.
\end{abstract}

\maketitle

\section{Equilibrating impurity and bath}

In this section we provide a more detailed analysis of our evolution scheme employed to prepare the equilibrium state, which is then used for the calculation of the Green's function. 
As explained in the main text, we expect the thermal conduction electrons to equilibrate with the impurity after bringing them into contact, hence converging towards the equilibrium state 
we are looking for. In practice, we decided to smoothly turn on the local and hybridization in a time-dependent fashion (see Fig.2 in main text), in order to minimize the creation of entanglement. 
In the following sections we investigate this equilibration procedure in different scenarios. 
We first consider the noninteracting case, where we show that the hybridization energy converges towards the correct thermal value, and Green's functions are reproduced correctly.
Afterwards, the effect of finite interaction $U$ is studied at zero temperature, showing that stronger interactions lead to slower equilibration, but accurate Green's functions can be obtained also for $U>0$. The last two sections are devoted to the analysis of finite temperature effects, and a summary including a more qualitative discussion of the equilibration process.

\begin{figure}[h]
	\centering
	\includegraphics[width=15cm]{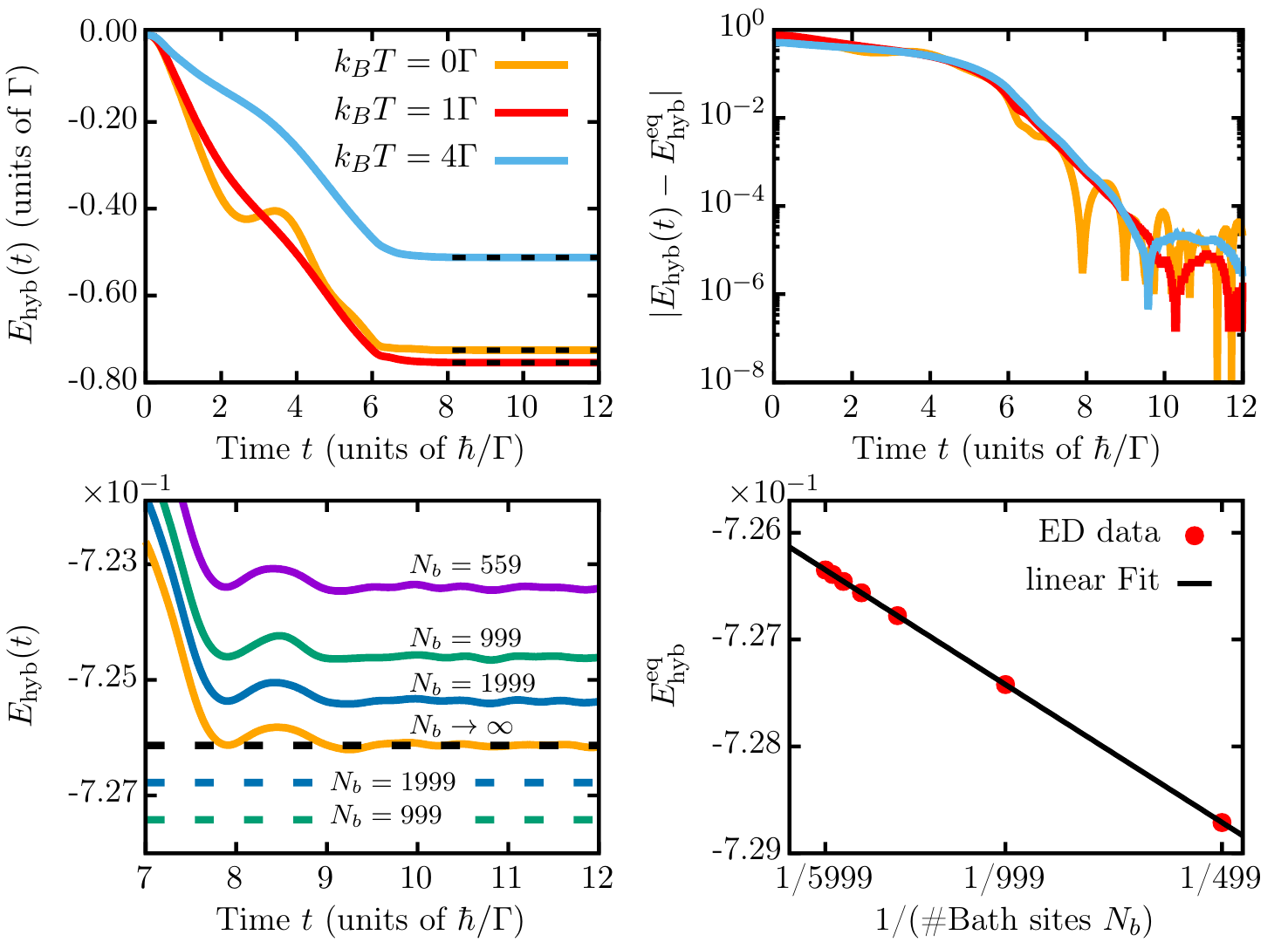}
	\caption{Analysis of the equilibration scheme for the noninteracting case $U=0$. Top left: Hybridization energy measured during equilibration with $\tau=12\hbar/\Gamma$. Dashed lines correspond to hybridization energy in the thermal state as obtained from exact diagonalization. Right: Difference between solid and dashed lines, showing excellent convergence of the hybridization energy. Bottom left: Size dependence of the hybridization energy at temperature $T=0$. Fermionic bath is linearly discretized into $N_b$ bath modes, and $N_b\rightarrow$ is obtained using orthogonal polynomials for the equilibration (solid lines) and through extrapolation of finite-size data in exact diagonalization (dashed lines). Right: Extrapolation of finite-size data  obtained from exact diagonalization.}
	\label{fig:NonintHybrEnergy}
\end{figure}

\subsection{Noninteracting case $U=0$}
At zero interaction, $U=0$, the single impurity Anderson model (SIAM) reduces to the resonant level model, and is solvable by means of exact diagonalization (ED), see Ref.~\cite{Serafini_book_2017}[(Appendix A)]. Hence, we are able to calculate expectation values not only in the ground state of the model, but also in the exact thermal state at any temperature $T$, to compare with our dynamical equilibration scheme. In the main text we show that the impurity occupation converges towards the thermal value, and here we would like to expand on this analysis. 
In particular, one quantity that we are able to compare is the hybridization energy, calculated as the expectation value 
$E_{\text{hyb}}(t)=\langle \psi(t)|\Ham_{\text{hyb}}|\psi(t)\rangle$ of the hybridization term
\begin{equation}
	\Ham_{\text{hyb}} = \sum_{\sigma} \sum_{k}  V_{k} \, \Big( \opddag{\sigma} \, \opc{k,\sigma} + \Hc \Big) 
	=\sum_{\sigma} \sum_{i=1}^2 J_{i,0} \left(\opddag{\sigma} \, \opa{i,0,\sigma} + \Hc \right),
\end{equation}
on the MPS state $|\psi(t)\rangle$ during equilibration. In \cref{fig:NonintHybrEnergy} (top left) we show the time-dependnce of the hybridization energy and the corresponding thermal state value obtained through ED. For all temperatures, we find very smooth convergence and excellent agreement at the end of the dynamics (see also \cref{fig:NonintHybrEnergy}, top right). 

At this point, let us briefly take a step away and discuss the effect of discretization. For ED we generally use linear discretization with different numbers of bath modes $N_b$; for MPS simulations we use either linear discretization (plus Lanczos tridiagonalization) or orthogonal polynomials. In \cref{fig:NonintHybrEnergy} (bottom left) we investigate the convergence of the hybridization energy for different discretizations at $T=0$, with linear discretization being used for all curves with finite $N_b$. In the limit $N_b\rightarrow \infty$ --- obtained using orthogonal polynomial discretization for the dynamics (orange curve) and extrapolation of finite-size ED data as presented in \cref{fig:NonintHybrEnergy} (bottom right) --- we observe excellent agreement between the equilibration scheme and ED. The finite-size behavior, however, is clearly different for ED and the MPS dynamics. This can be understood from the following arguments: For the MPS dynamics, only chain sites sufficiently close to the impurity do have an impact on the hybridization energy, since any excitation created due to the interaction of impurity and fermion bath --- being either a particle in the empty chain or a hole in the filled chain --- is traveling at finite speed along the chain. Hence, it cannot reach chain sites too far away from the impurity by the end of the dynamics. On the other hand, in equilibrium simulations (DMRG or ED)  {\textit{all}} chain sites contribute to the equilibrium state, including those far away from the impurity and unreachable by the dynamics. The discrepancy at finite bath size is therefore expected, and, in all of the energy comparisons, we consider the limit $N_b\rightarrow \infty$. 

\begin{figure}[h]
	\centering
	\includegraphics[width=15cm]{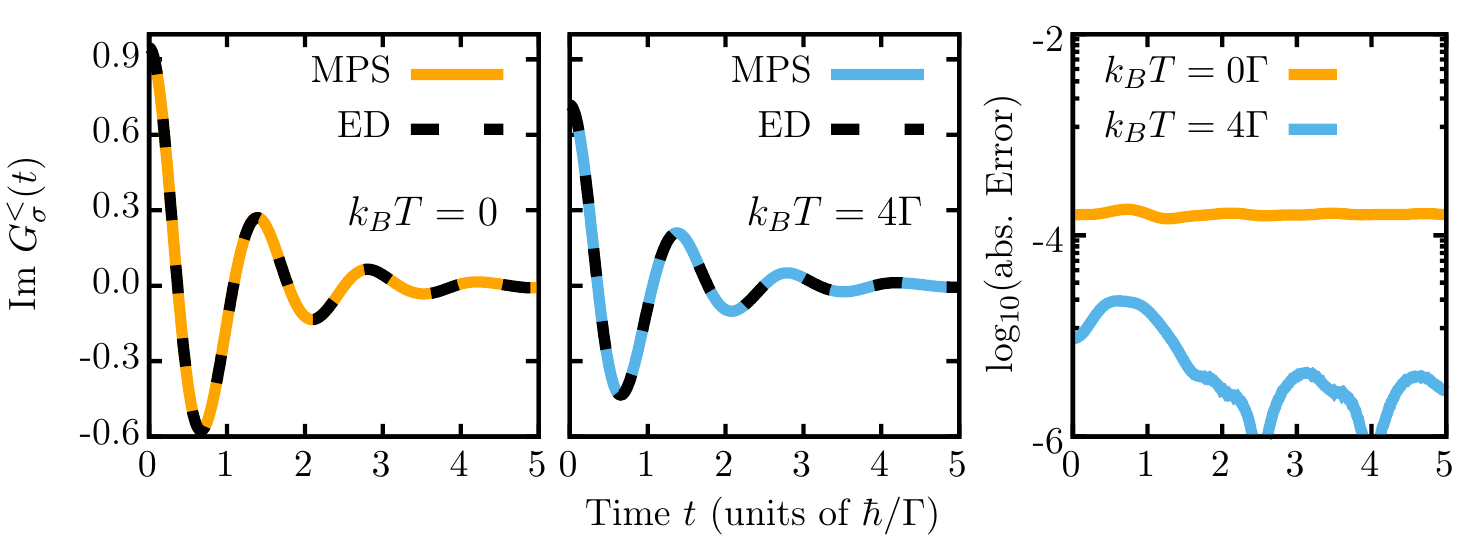}
	\caption{Left and middle: Imaginary part of the lesser Green's function for $U=0$. Initial states to calculate $G^{<}_{\sigma}$ were prepared through equilibration with $\tau=12\hbar/\Gamma$ (solid line) and exact diagonalization (dashed line). Right: Absolute error of the lesser Green's function when using real-time equilibration as compared to ED results.}
	\label{fig:NonintGF}
\end{figure}

Let us turn to the object of interest: the Green's function. For the comparison with ED, we focus on the lesser Green's function 
$G^{<}_{\sigma}(t) = i\,\Tr_{\sys}( \opddag{\sigma}\opd{\sigma}(t)\hat{\rho}_{\sys})$, since in contrast to the retarded Green's function it {\em does} show temperature dependence even 
at $U=0$. We compare the Green's function obtained from ED, starting from the exact thermal state, with the one calculated with our MPS-based approach, where a real-time equilibration is used to prepare the initial state. As shown in \cref{fig:NonintGF}, we find excellent agreement, verifying that our approach correctly reproduces the temperature dependence of the Green's function.

\begin{figure}[h]
	\centering
	\includegraphics[width=18cm]{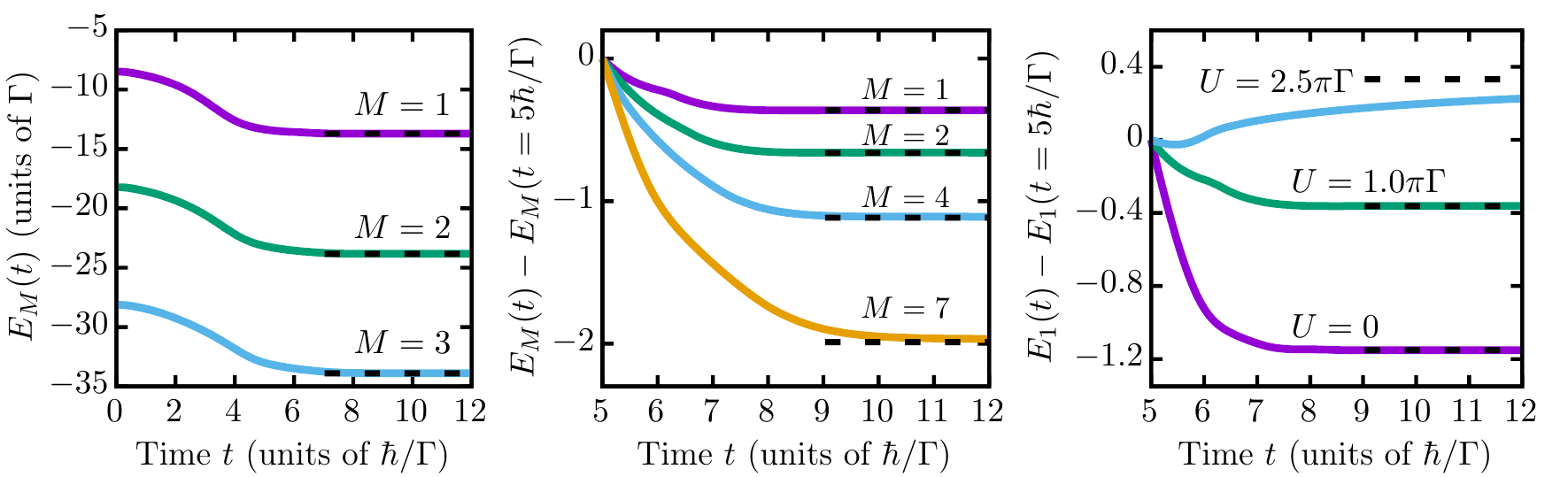}
	\caption{Analysis of the equilibration scheme at fixed temperature $T=0$. Left: Dynamics of the partial energy for different $M$ at fixed interaction $U=\pi\Gamma$ and $\tau=12\hbar/\Gamma$. Dashed lines indicate ground state values extrapolated to the thermodynamic limit $N_b\rightarrow \infty$. Middle: Convergence behaviour of partial energies. We subtract $E_M(t=5\hbar/\Gamma)$ to have the same energy scale for all curves even with different $M$. Right: Convergence of the partial energy $E_1$ for different interactions $U$. Stronger repulsion leads to slower equilibration.}
	\label{fig:InteractingT0Energies}
\end{figure}

\begin{figure}[h]
	\centering
	\includegraphics[width=18cm]{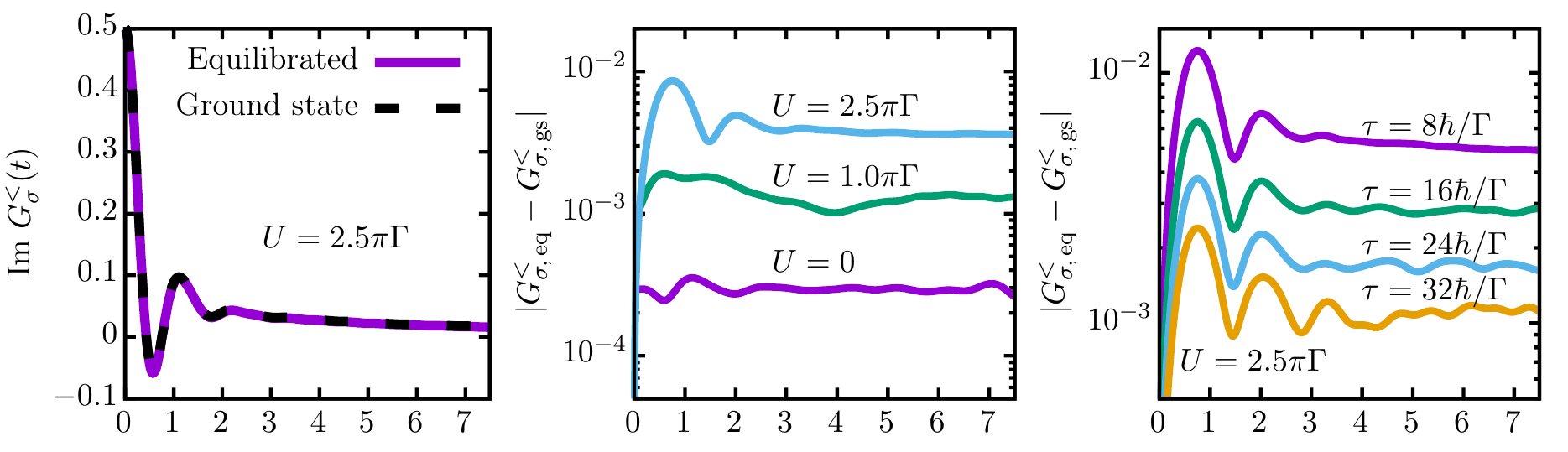}
	\caption{Left: Imaginary part of the lesser Green's function for $U=2.5\pi\Gamma$ using the initial state obtained from equilibration (with $\tau=32\hbar/\Gamma$) and DMRG ground state search. Middle: Error of the lesser Green's function obtained after equilibrating for $\tau=12\hbar/\Gamma$ as compared to starting from the DMRG ground state. Stronger repulsion leads to slower equilibration, hence to larger errors in the Green's function. Right: Error of $G^{<}_{\sigma}$ for different equilibration times at $U=2.5\pi\Gamma$, which is found to decreases smoothly as we let the system equilibrate longer.}
	\label{fig:InteractingT0GF}
\end{figure}

\subsection{Interacting case at zero temperature}
At $T=0$ and finite interaction $U$ we can benchmark the equilibration procedure with the ground state as obtained from DMRG calculations. 
We concentrate on the Green's function as well as the partial energy, taking into account only certain terms of the Hamiltonian. The full SIAM with truncated chain length (finite $L$) is given by 
\begin{eqnarray}  \label{eq:HamChain}
\Ham_{\SIAM,L} = \Ham_{\text{loc}}  + \Ham_{\text{hyb}} +  \sum_{\sigma} \sum_{i=1}^2 \bigg(
\sum_{n=1}^{L-1}  \left( J_{i,n} \opadag{i,n,\sigma}\opa{i,n-1,\sigma} + \Hc \right)  
+ \sum_{n=0}^{L-1}  E_{i,n} \, \opadag{i,n,\sigma}\opa{i,n,\sigma} \bigg) \;. 
\end{eqnarray}
We define the partial energies to be expectation values $E_{M}=\langle \psi(t)|\Ham_{\SIAM,M}|\psi(t)\rangle$ of this operator, with $M\leq L$, such that
only chain sites closest to the impurity are considered for the energy measurement. As expected, we find the partial energies to converge: for fixed $U=\pi\Gamma$ the final values are in perfect agreement with the corresponding expectation value in the DMRG ground state (see \cref{fig:InteractingT0Energies}, left and middle panel). Similarly to the previous section, static DMRG data were obtained using linear discretization with various numbers of bath modes $N_b$, and extrapolated to the thermodynamic limit employing a polynomial fitting function $E_M(N_b)=a (N_b)^{-c}+d$, with fitting parameters $a,c,d$. Zooming into the convergence behavior  at the end of the equilibration (that is why we subtract $E_M(t=5\hbar/\Gamma)$), we find that smaller $M$ partial energies converge faster, implying that bath sites closer to the impurity equilibrate faster than sites further away. Investigating the effect of the interaction $U$, we find that stronger interactions lead to significantly slower equilibration (see \cref{fig:InteractingT0Energies}, right panel). Hence, longer evolution times are required to reach equilibrium. This behavior also impacts the accuracy of the Green's function \cref{fig:InteractingT0GF} (left panel). Comparing the Green's function obtained after the equilibration step with the one computed starting from the DMRG ground state, we observe larger deviations as $U$ increases (see \cref{fig:InteractingT0GF}, middle panel). Indeed this error is due to insufficient equilibration and can be controlled through the equilibration time $\tau$ \cref{fig:InteractingT0GF} (right panel). The longer the equilibration, the more accurate the Green's function. In practice, it is therefore important to converge the Green's function with respect to $\tau$.

\subsection{Effect of finite temperature at $U=2.5\pi\Gamma$}
\begin{figure}[h]
	\centering
	\includegraphics[width=8cm]{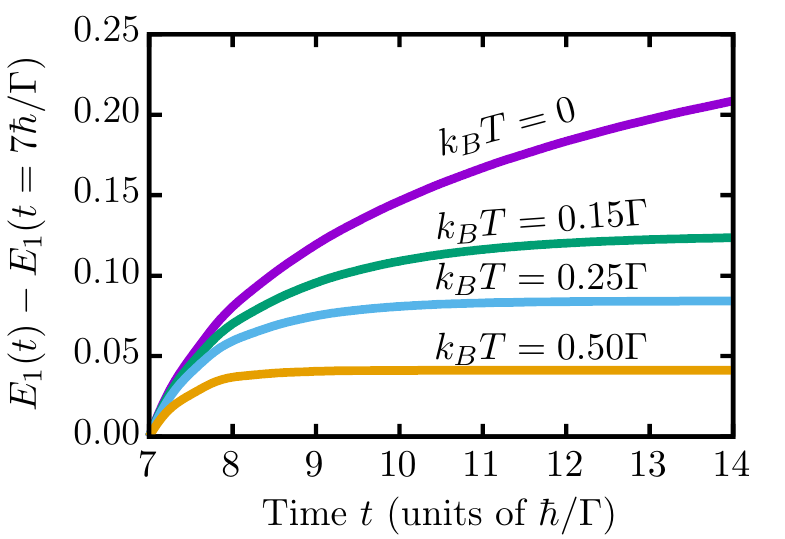}
	\caption{Convergence of the partial energy $E_1(t)$ for fixed interaction $U=2.5\pi\Gamma$ and equilibration time $\tau=14\hbar/\Gamma$ at various temperatures. Higher temperature leads to significantly faster convergence.}
	\label{fig:InteractingFiniteTEnergy}
\end{figure}

In this section we investigate the effect of temperature on the equilibration in the interacting case. We fix the interaction to be $U=-2\varepsilon_d=2.5\pi\Gamma$, where at $T=0$ we observed rather slow convergence of the partial energy and the Green's function with respect to the total equilibration time $\tau$. We start analyzing the convergence behavior of the partial energy $E_1(t)$ at fixed $\tau=14\hbar/\Gamma$ in \cref{fig:InteractingFiniteTEnergy}. While at $T=0$ it clearly has not converged until the end of dynamics, it does converge significantly faster at higher temperatures. These findings make us hope that also the Green's function will converge much faster at finite temperature, so let us have a look at it.

\begin{figure}[h]
	\centering
	\includegraphics[width=18cm]{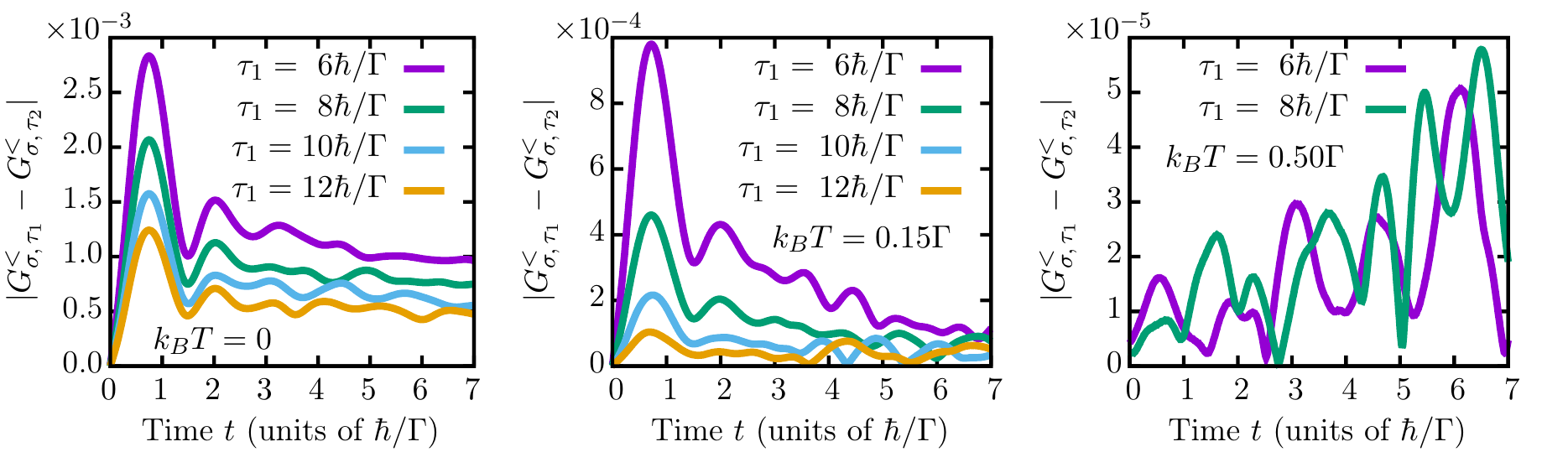}
	\caption{Convergence of the Green's function is presented for temperatures $k_BT=0$ (left), $k_BT=0.15\Gamma$ (middle) and $k_BT=0.5\Gamma$ (right). Plots show the difference of two Green functions obtained after  equilibrating for total time $\tau_1$ and $\tau_2=\tau_1+2\hbar/\Gamma$. Higher temperature leads to significantly reduced differences, indicating faster convergence of the equilibration with respect to $\tau$.}
	\label{fig:InteractingFiniteTGF}
\end{figure}

At both finite interaction and temperature -- in contrast to the previous sections -- we do not have any reference results available we could compare the Green's function to. For this reason we use an alternative strategy to investigate convergence with $\tau$: We calculate the difference $|G^{<}_{\sigma,\,\tau_1}-G^{<}_{\sigma,\,\tau_2}|$ of two Green's functions obtained after equilibration with different equilibration times $\tau_1$ and $\tau_2$, where we fix $\tau_2=\tau_1+2\hbar/\Gamma$. At convergence this difference should vanish, since the Green's functions become independent of the equilibration time. In \cref{fig:InteractingFiniteTGF} we show this convergence measure for zero (left panel) and finite temperature (middle and right panel), and observe that for $T=0$ it decreases as we increase $\tau_1$ (and implicitly $\tau_2$) as expected. At $T>0$ we find the Green's function to converge significantly faster as compared to the zero temperature case (note the different scale of the y-axis). While at $k_BT=0.15\Gamma$ our difference measure nicely decreases down to only $10^-4$, for $k_BT=0.5\Gamma$ it already stays below that value even for short equilibration with $\tau_1=6\hbar/\Gamma$.

\subsection{Summary and discussion}
We have studied in detail the procedure we use to prepare the equilibrium state at finite temperature. We find that -- after turning on the hybridization (and the local term) -- impurity occupation, hybridization energy, and more generally, partial energies taking into account only several chain sites closest to the impurity, converge towards their respective values in the exact equilibrium state. In particular, we focused on the cases of either zero interaction or zero temperature, where exact results for the equilibrium state are available, but we expect our conclusion to hold at finite interaction and temperature as well. We have seen that equilibration takes more and more time as the interaction $U$ increases. On the other hand, temperature tends to speed up the process. Those findings might be related to the Kondo peak, which becomes more narrow at stronger interaction, but decreases in height as temperature increases. 

The qualitative picture we have in mind for the equilibration is that -- in the chain geometry -- fermionic sites locally converge towards their equilibrium state, with sites close to the impurity converging faster than sites further away. With this picture, we can discuss the issue of unitary dynamics versus equilibration: Since we consider a closed quantum system (consisting of impurity and bath fermions) with Hamiltonian dynamics only, the evolution is unitary. Hence, the quantum state contains information about the initial state at all times. At first this might seem to be in contradiction to the idea of converging towards some equilibrium steady state. However, excitations -- created in the chain sites connected to the impurity while/after turning on the hybridization -- are able to travel along the (infinitely long) chain forever. Hence, they can carry information and residual energy of the initial state away from the impurity, while sites closer to the impurity are able to equilibrate. It is important to note that the calculation of the Green's function requires the application of creation/ annihilation operators at the impurity, which does not interfere with the non-equilibrated part of the state far away from the impurity. For this reason we are able to obtain accurate Green's functions that do not contain any signature of the equilibration procedure.

\section{Parameters and numerical details}
In our tensor network simulations we have two crucial parameters to restrict the number of states kept in the MPS: 1) first, we have the truncated weight, corresponding to the summed probability 
of discarded states; 2) second, we also use a hard cutoff on the number of states we keep, the so-called {\em bond dimension}. 
We use a small truncated weights of $w_t=10^{-12}$ and a maximum bond dimension of typically $D=150$. 
Only for the zero temperature simulations with the standard chain mapping, the bond dimension was increased to $D=400$, in order to deal with the increase of entanglement. 
For all real time evolutions we use time steps in the range $\Delta t=0.01-0.02\hbar/\Gamma$. 
(The largest energy scale in our system is given by the half bandwidth $W=10\Gamma$.) 
We verified that our results are converged in all relevant numerical parameters. 
We further found that setting a minimum bond dimension of $D_{\text{min}}\approxeq 20$ (i.e., keeping even some states with low probability) can be beneficial for the numerical stability of TDVP when dealing with next-nearest neighbor interactions.
We confirmed that 2-site TDVP delivers accurate results in this scenario by comparison of different MPS orderings. We compared our results with simulations employing the two separated chains ordering (see Fig.\textcolor{blue}{1}(d) in the main text). Here, projection errors are absent due to the Hamiltonian terms being at most nearest neighbor in distance \cite{AoP_Paeckel_2019}. The 'separated chains' ordering works fine at low temperatures, where we carried out the comparison, but shows strong entanglement growth for higher temperatures.
A more detailed analysis of these technical details, however, will be presented elsewhere.
In our simulations we explicitly exploit conservation of particle number and spin to speed up the calculations~\cite{PRA_Singh_2010,PRB_Singh_2011,Scipost_Silvi_2019}.
We have seen that equilibration is faster for higher temperature. For this reason, we use different equilibration times $\tau=12\hbar/\Gamma$ ($k_BT=0.15\Gamma$), $\tau=8\hbar/\Gamma$ ($k_BT=0.25\Gamma$), $\tau=6\hbar/\Gamma$ ($k_BT=0.5\Gamma$), and $\tau=4\hbar/\Gamma$ ($k_BT=\Gamma$ and $k_BT=4\Gamma$). 
Green's functions are computed up to time $t_f=7.5\hbar/\Gamma$ (finite $T$), $t_f=30\hbar/\Gamma$ ($T=0, U=2.5\pi\Gamma$) and $t_f=40\hbar/\Gamma$ ($T=0, U=3.25\pi\Gamma$) before applying linear prediction to extrapolate the exponential tail~\cite{PRB_White_2008,PRB_Schollwoeck_2009,PRB_Verstraete_2015}. 

Finally, let us discuss the discretization issue: In the dynamical case, which employs the equilibration procedure, we are able to work directly in the continuum limit $N_b\rightarrow \infty$ 
using orthogonal polynomials, as discussed previously. In practice, we truncate the infinitely long chain such that no excitation --- being either a particle in the empty chain or a hole in the filled chain --- does reach the end of the chain during the dynamics~\cite{PRB_Schroeder_2016}. 
Clearly, this does not introduce any error since the truncation does not have any impact on the quantum state at any time (as long as the chain is long enough). 
At $T=0$ instead, where DMRG can be used to calculate the equilibrium (ground state) state, we use linear discretization of the bath. 
This turned out to yield more accurate results for the ground state. For the calculations of Green's functions at $T=0$ we linearly discretized the bath into $N_b=559$ modes, 
mapping 280 modes to the empty chain and 279 modes to the filled chain in our approach. We observed convergence with respect to $N_b$.

\subsection{Performance dependence on bath size}
\begin{figure}[t]
	\centering
	\includegraphics[width=8cm]{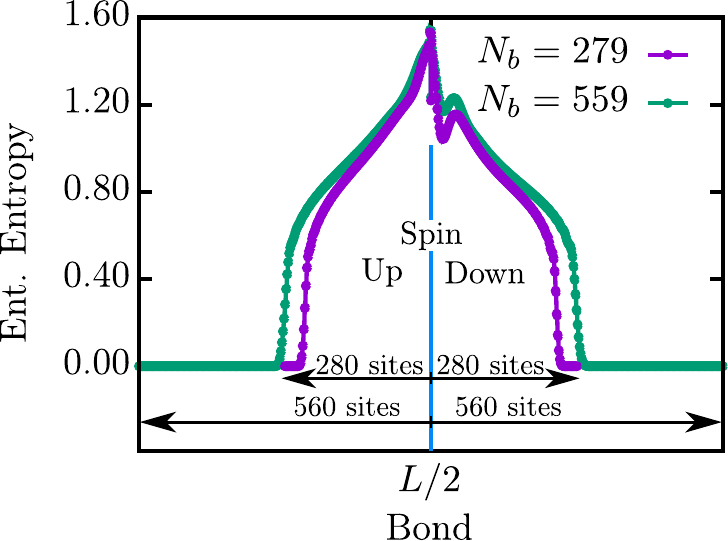}
	\caption{Entanglement entropy along the MPS at our final time $t=30\hbar/\Gamma$ at zero temperature, for the symmetric SIAM. At $T=0$ we separate spin-up and spin-down degrees of freedom into two chains, connected in the middle (see also \cref{fig:EntStar}). Two different bath sizes $N_b=279$ and $N_b=559$ are considered, where $L/2$ labels the middle of the MPS. Note that the curves are centered around $L/2$ such that the smaller $N_b$ curve does not extend to the boundaries of the window. During time evolution the nonzero entanglement region grows  continuously (see \cref{fig:EntStar}) and almost reaches the end of the $N_b=279$ chains. Hence, any further evolution with this discretization would lead to significant errors in the dynamics.}
	\label{fig:EntanglementdifSizes}
\end{figure}

For practical applications it can be crucial to have a fine discretization of the bath, in order to sample the hybridization function as accurate as possible. However, fine discretization translates into a large bath with many discrete conduction band levels. For an algorithm to be efficient we require it to scale well with the number of bath modes. We investigate this issue by analyzing the entanglement entropy at the end of our zero temperature Green's function calculation (see \cref{fig:EntanglementdifSizes}). We find that the number of bath sites -- translating into the equivalent number of chain sites -- barely affects the entanglement structure in the MPS. Within the region reached by excitations (nonzero entanglement) differences between the two curves are negligible. The larger bath has an extended region not reached by any excitation, and, since the state has zero entanglement here, time evolution is very efficient here. Hence, our approach shows excellent scaling with the number of sites and allows to simulate accurately discretized baths without any significant increase of computational costs.

\subsection{Comparison with Star geometry}
\begin{figure}[t]
	\centering
	\includegraphics[width=15cm]{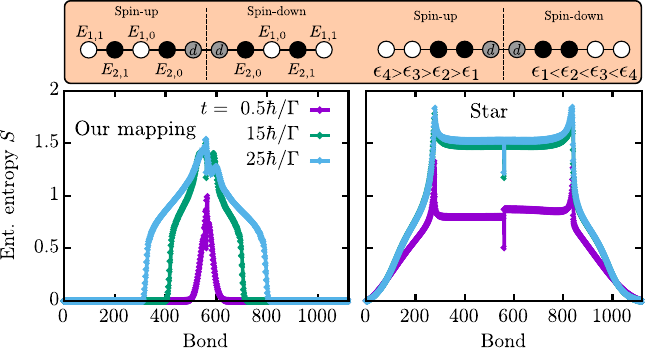}
	\caption{Top: Ordering of the sites in the MPS for our chain mapping based method (left) and the star geometry (right). At $T=0$ we split spin degrees of freedom into two chains, due to the reduced amount of entanglement. Grey symbols represent impurity sites, while black and white symbols correspond to modes that are filled and empty (in absence of hybridization), respectively. Bottom: Entanglement entropy along the MPS during the caluclation of the greater Green's function at $T=0$ and different times $t$, for the symmetric SIAM.}
	\label{fig:EntStar}
\end{figure}

In the main text we already discussed the entanglement entropy in our method and compared it with the original chain mapping, finding significantly lower entanglement in our approach. However, it has been shown by Wolf \textit{et~al.}\cite{Wolf_PRB14} that the original chain mapping based approach is rather inefficient and a direct simulation in the star geometry can reduce the entanglement in the MPS. For our comparison we employ the MPS ordering of Ref.\cite{PRX_Bauernfeind_2017}, which includes an additional splitting of spin-up and spin-down as compared to Ref.~\cite{Wolf_PRB14}. We found this splitting of spin degrees of freedom to lower the entanglement in our scenario considered here. Hence, the star geometry MPS consists of two impurity sites in the middle, one for spin-up and one for spin-down (see \cref{fig:EntStar}). Both impurity sites are coupled to their individual conduction modes, represented in the star geometry. Conduction modes are ordered according to their energy, starting with the lowest energy mode next to the corresponding impurity, and increasing in energy as we move away from the impurity sites. For simplicity we restrict our analysis to $T=0$, where neither the thermofield transformation nor the equilibration process are needed. For the star geometry find the entanglement to show a plateau in the middle of the MPS (see \cref{fig:EntStar}), with almost constant entanglement. We believe that this plateau is due to the entanglement between the impurity and the conduction modes close to the Fermi energy. These modes, however, are located in the middle of each bath, where we also find entanglement peaks (one for spin-up and one for spin-down). In our chain mapping based method we find clearly lower entanglement along the MPS, leading to much faster simulations. However, the entanglement structure highly depends on the ordering of the sites in the MPS, especially in the star geometry, and different orderings might reduce the entanglement. Here, the main disadvantage of the star geometry seems to be the spacial separation of the impurity and modes close to the Fermi energy, leading to significantly worse scaling with the number of bath sites as compared to our chain mapping based approach. This might be avoided moving to a nonlinear tensor network \onlinecite{PRL_Rams_2020}. We further note that we have concentrated on the Kondo regime here, and that the entanglement can strongly depend on the physical parameters. Hence, further analysis is needed on these issues.

\subsection{Convergence with bond dimension $D$}
\begin{figure}[h]
	\centering
	\includegraphics[width=16cm]{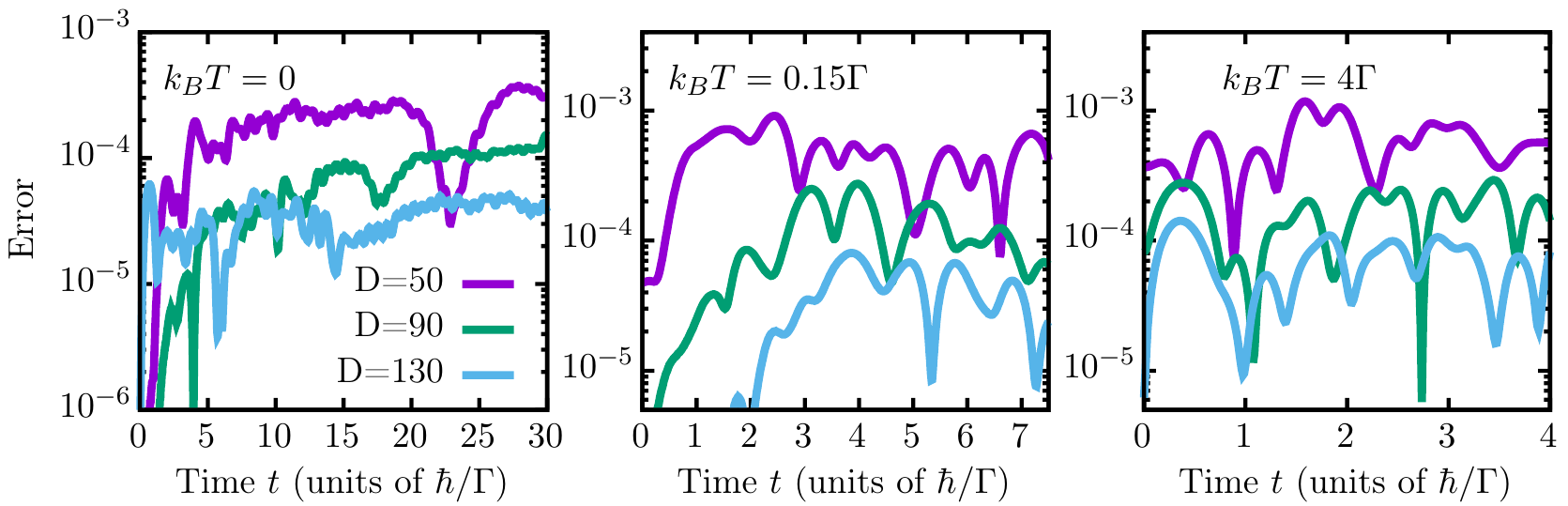}
	\caption{Error $|G^{<}_{\text{ref}}(t)-G^{<}_{\text{D}}(t)|$ in the lesser Green's function as compared to the reference simulation $(D=250, w_t=10^{-14})$, for different bond dimensions $D$ ($w_t=10^{-12}$). The symmetric SIAM is considered, with $\varepsilon_d=-1.25\pi\Gamma$ and $U=2.5\pi\Gamma$, and temperatures $k_BT=0$ (left), $k_BT=0.15\Gamma$(middle) and $k_BT=4\Gamma$ (right). We consider different maximum simulation times depending on temperature due to the faster decay of the Green's function at higher temperature.}
	\label{fig:BondGFconvergence}
\end{figure}

In tensor network algorithms the bond dimension $D$ is the key numerical parameter, defining how many states are kept at each bond, hence determining the accuracy of the simulation. For an algorithm to be efficient, it is crucial that the results converge sufficiently fast with respect to the bond dimension. Since no exact results are available, we investigate the convergence of the Greens function as compared to a reference simulation with large bond dimension $D=250$ and small truncated weight $w_t=10^{-14}$. We define the error to be the difference $|G^{<}_{\text{ref}}(t)-G^{<}_{\text{D}}(t)|$ between the reference simulation and simulations with bond dimension $D$ and the usual truncated weight $w_t=10^{-12}$.  In \cref{fig:BondGFconvergence} we show the convergence of the lesser Green's function with respect to the bond dimension at different temperatures. We find already qualitatively good results at bond dimensions as low as $D\sim50$, improving further when increasing $D$ as expected. For our simulations in the main paper we used $D=150$ to reduce numerical errors to a minimum. 

